\documentclass{PoS}
\pdfoutput=1
\usepackage{amsfonts, amsmath, amsthm, amssymb}
\usepackage{graphicx}
\usepackage[labelformat=empty]{subfig}
\newcommand{\dfourx}{\mathrm{d}^4 x}

\newcommand{\quarkthreed}{\langle\overline{q}q\rangle}
\newcommand{\gluonfourd}{\langle\alpha G^2\rangle}
\newcommand{\mixed}{\langle g\overline{q} \sigma G q\rangle}
\newcommand{\gluonsixd}{\langle g^3 G^3\rangle}
\newcommand{\vev}[1]{\langle\Omega|#1|\Omega\rangle}
\newcommand{\angled}[1]{\langle #1\rangle}

\newcommand{\mev}{\ensuremath{\text{MeV}}}

\title{Mass Predictions of Open-Flavour Hybrid Mesons from QCD Sum Rules}

\ShortTitle{Mass Predictions of Open-Flavour Hybrid Mesons from QCD Sum Rules}

\author{\speaker{Jason Ho}\\
         Department of Physics \& Engineering Physics\\
        University of Saskatchewan\\
        E-mail: \email{j.ho@usask.ca}}

\author{Derek Harnett\\
		Department of Physics\\
        University of the Fraser Valley\\
        E-mail: \email{derek.harnett@ufv.ca}}
        
\author{Tom Steele\\
		Department of Physics \& Engineering Physics\\
        University of Saskatchewan\\
        E-mail: \email{tom.steele@usask.ca}}

\abstract{Within QCD, colourless states may be constructed corresponding to exotic matter outside of the traditional quark model. Experiments have recently observed tetraquark and pentaquark states, but no definitive hybrid meson signals have been observed. With the construction of the PANDA experiment at FAIR, and with full commissioning of the GlueX experiment at JLab expected to be completed this year, the opportunity for the observation of hybrid mesons has greatly increased. However, theoretical calculations are necessary to ascertain the identity of any experimental resonances that may be observed. We present selected QCD sum rule results from a full range of quantum numbers for open-flavour hybrid mesons with heavy valence quark content, including non-perturbative condensate contributions up to six-dimensions.}

\FullConference{38th International Conference on High Energy Physics\\
                  3-10 August 2016\\
                 Chicago, USA}

\begin{document}

\section{Correlation Functions of Open-Flavour Hybrid Mesons}
The seminal application of QCD sum-rules to heavy-light hybrids was performed by 
Govaerts, Reinders, and Weyers~\cite{GovaertsReindersWeyers1985} (hereafter GRW).
Therein, they considered four distinct currents covering $J\in\{0,\,1\}$
in an effort to compute a comprehensive collection of ground state hybrid masses.
For all heavy-light hybrids, the square of the ground state hybrid mass was
uncomfortably close to the continuum threshold (with a typical separation
of roughly 10--15~$\mev$), and it was noted that even a modest hadron width
would result in the resonance essentially merging with the continuum.

In this paper, we briefly review \cite{HoHarnettSteele2016} where we extended the work of GRW \cite{GovaertsReindersWeyers1985} by  
including both 5d mixed and 6d gluon condensate contributions in
our correlator calculations.
As noted in GRW, for heavy-light hybrids, condensates involving light quarks
are multiplied by a heavy quark mass allowing for the possibility
of a numerically significant contribution to the correlator and to the sum-rules.  
By this reasoning, the 5d mixed condensate could be a significant component 
of a QCD sum-rules application to the hybrid systems under consideration.
As well, recent sum-rules analyses of closed, 
heavy hybrids~\cite{ChenKleivSteeleEtAl2013} have demonstrated that the 6d gluon condensate can have an important stabilizing
effect on what were, in the pioneering work~\cite{GovaertsReindersWeyers1985,GovaertsReindersRubinsteinEtAl1985,GovaertsReindersFranckenEtAl1987}, unstable analyses.

Following GRW, we define open hybrid interpolating currents, $j_{\mu}=\frac{g_s}{2} ~ \overline{Q} ~ \Gamma^{\rho} \lambda^a q~  \mathcal{G}^a_{\mu\rho}$,
where $g_s$ is the strong coupling and $\lambda^a$ are the Gell-Mann matrices.
The field $Q$ represents a heavy charm or bottom quark with mass $M_Q$
whereas $q$ represents a light up, down, or strange quark with mass $m_q$. 
The Dirac matrix $\Gamma^{\rho}$ satisfies $\Gamma^{\rho} \in \{\gamma^{\rho},\,\gamma^{\rho}\gamma_5\}$
and the tensor $\mathcal{G}^a_{\mu\rho}$, the portion of $j_{\mu}$ containing 
the gluonic degrees of freedom, satisfies $\mathcal{G}^a_{\mu\rho} \in \{G^a_{\mu\rho},\,\tilde{G}^a_{\mu\rho} = \frac{1}{2}\epsilon_{\mu\rho\nu\sigma}G^a_{\nu\sigma}\}$
where $G^a_{\mu\rho}$ is the gluon field strength and $\tilde{G}^a_{\mu\rho}$ is its dual defined using the totally antisymmetric Levi-Civita 
symbol $\epsilon_{\mu\rho\nu\sigma}$.

For each of the four currents defined through $j_{\mu}$ above,
we consider a corresponding diagonal, two-point correlation function
\begin{equation}
\label{correlator}
  \Pi_{\mu\nu}(q) = i\int\dfourx e^{i q\cdot x} 
    \vev{\tau j_{\mu}(x)j^{\dag}_{\nu}(0)}
  = \frac{q_{\mu}q_{\nu}}{q^2}\Pi^{(0)}(q^2) 
   + \left(\frac{q_{\mu}q_{\nu}}{q^2}-g_{\mu\nu}\right)\Pi^{(1)}(q^2).
\end{equation}
The tensor decomposition in~(\ref{correlator}) is such that $\Pi^{(0)}$ probes
spin-0 states while $\Pi^{(1)}$ probes spin-1 states.
We will reference each of the $\Pi^{(0)}$ and $\Pi^{(1)}$
according to the $J^{PC}$ combination it would have were we investigating closed 
rather than open hybrids; however, to stress that the $C$-value can not be taken 
literally, we will enclose it in brackets.
\begin{center}
\captionof{table}{The $J^{P(C)}$ combinations probed through different choices of $\Gamma^{\rho}$ and $\mathcal{G}^a_{\mu\rho}$.}
\begin{tabular}[h]{c|c||c}
  $\Gamma^{\rho}$ & $\mathcal{G}^a_{\mu\rho}$ & $J^{P(C)}$\\
  \hline
  $\gamma^{\rho}$ & $G^a_{\mu\rho}$ & $0^{+(+)},\,1^{-(+)}$ \\
  $\gamma^{\rho}$ & $\tilde{G}^a_{\mu\rho}$ & $0^{-(+)},\,1^{+(+)}$ \\
  $\gamma^{\rho}\gamma_5$ & $G^a_{\mu\rho}$ & $0^{-(-)},\,1^{+(-)}$ \\
  $\gamma^{\rho}\gamma_5$ & $\tilde{G}^a_{\mu\rho}$ & $0^{+(-)},\,1^{-(-)}$
\label{JPC_table}
\end{tabular}
\end{center}

We calculate the correlators~(\ref{correlator}) within the operator product expansion (OPE)
in which perturbation theory is supplemented by a collection of non-perturbative terms,
each of which is the product of a perturbatively computed Wilson coefficient 
and a non-zero vacuum expectation value (VEV) or condensate.  
We include condensates up to and including those of dimension (d) six:
\begin{gather}
  \quarkthreed=\angled{\overline{q}_i^{\alpha} q_i^{\alpha}}
    \label{condensate_quark_three}\\
  \gluonfourd=\angled{\alpha_s G^a_{\mu\nu} G^a_{\mu\nu}}\label{condensate_gluon_four}\\
  \mixed=\angled{g_s \overline{q}_i^{\alpha}\sigma^{\mu\nu}_{ij}
    \lambda^a_{\alpha\beta} G^a_{\mu\nu} q_j^{\beta}}
    \label{condensate_mixed}\\
  \gluonsixd=\angled{g_s^3 f^{abc} G^a_{\mu\nu}G^b_{\nu\rho}G^c_{\rho\mu}}.\label{condensate_gluon_six}
\end{gather}
where the VEVs~(\ref{condensate_quark_three})--(\ref{condensate_gluon_six}) 
are respectively referred to as
the 3d~quark condensate,
the 4d~gluon condensate,
the 5d~mixed condensate, and
the 6d~gluon condensate.

\begin{figure}[ht!]
\subfloat[]{%
    \includegraphics[width=.142\textwidth]{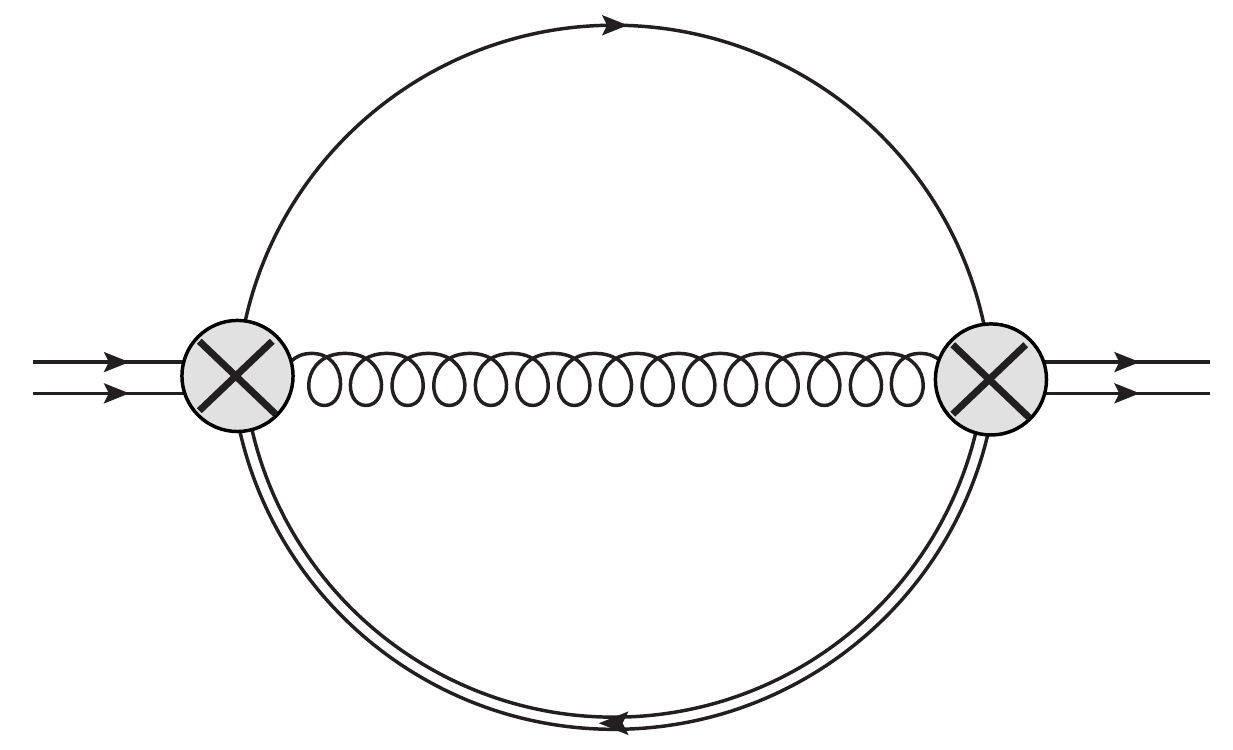}}\hfill
  \subfloat[]{%
    \includegraphics[width=.142\textwidth]{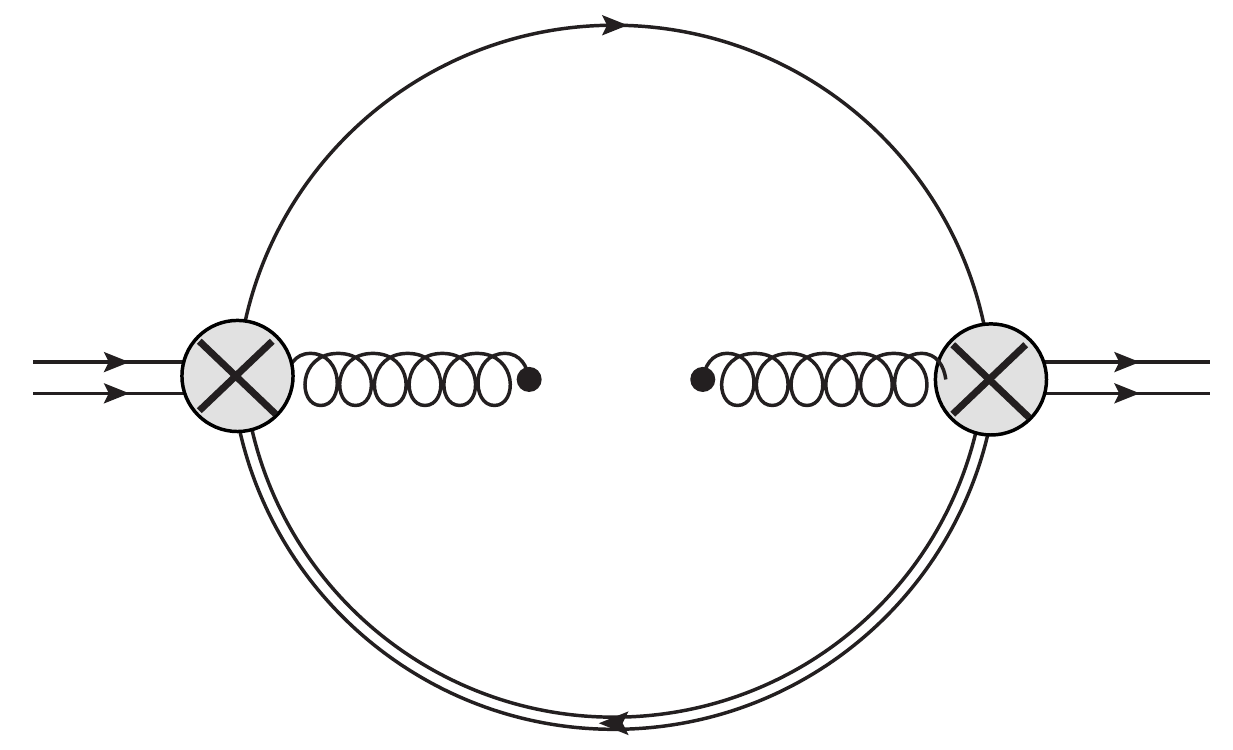}}\hfill
  \subfloat[]{%
    \includegraphics[width=.142\textwidth]{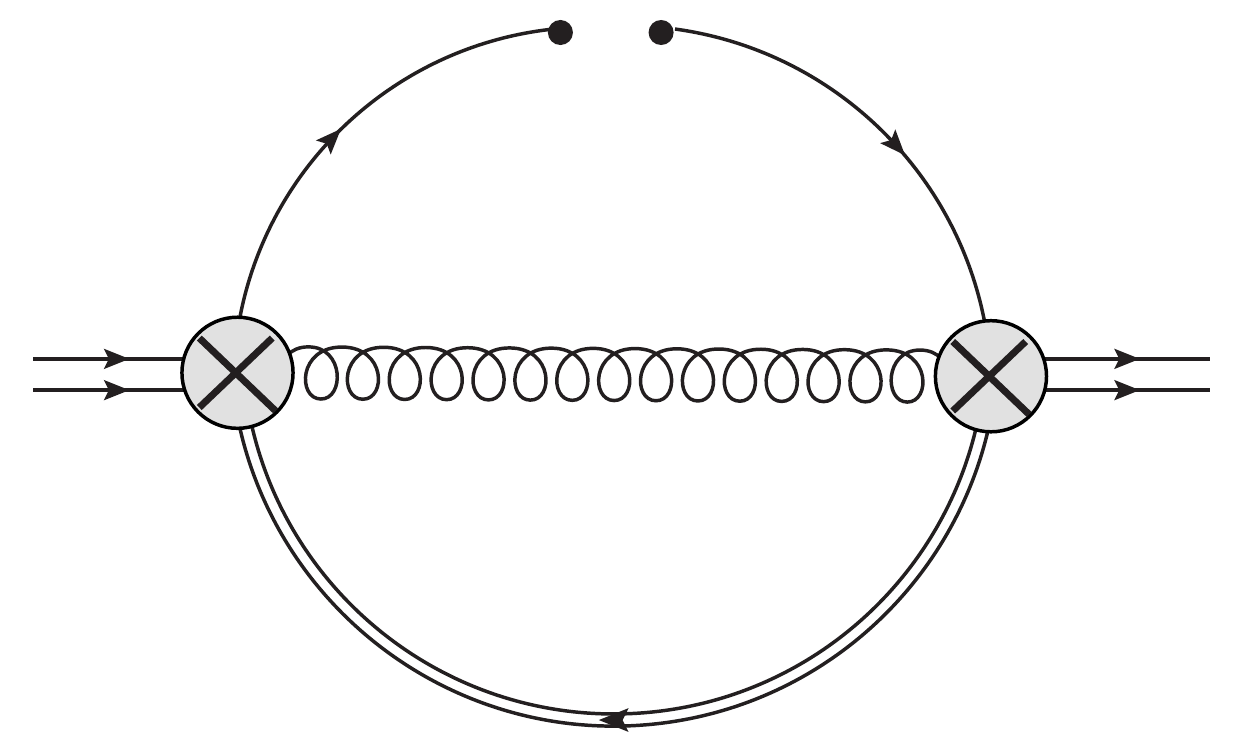}}\hfill
  \subfloat[]{%
  	\includegraphics[width=.142\textwidth]{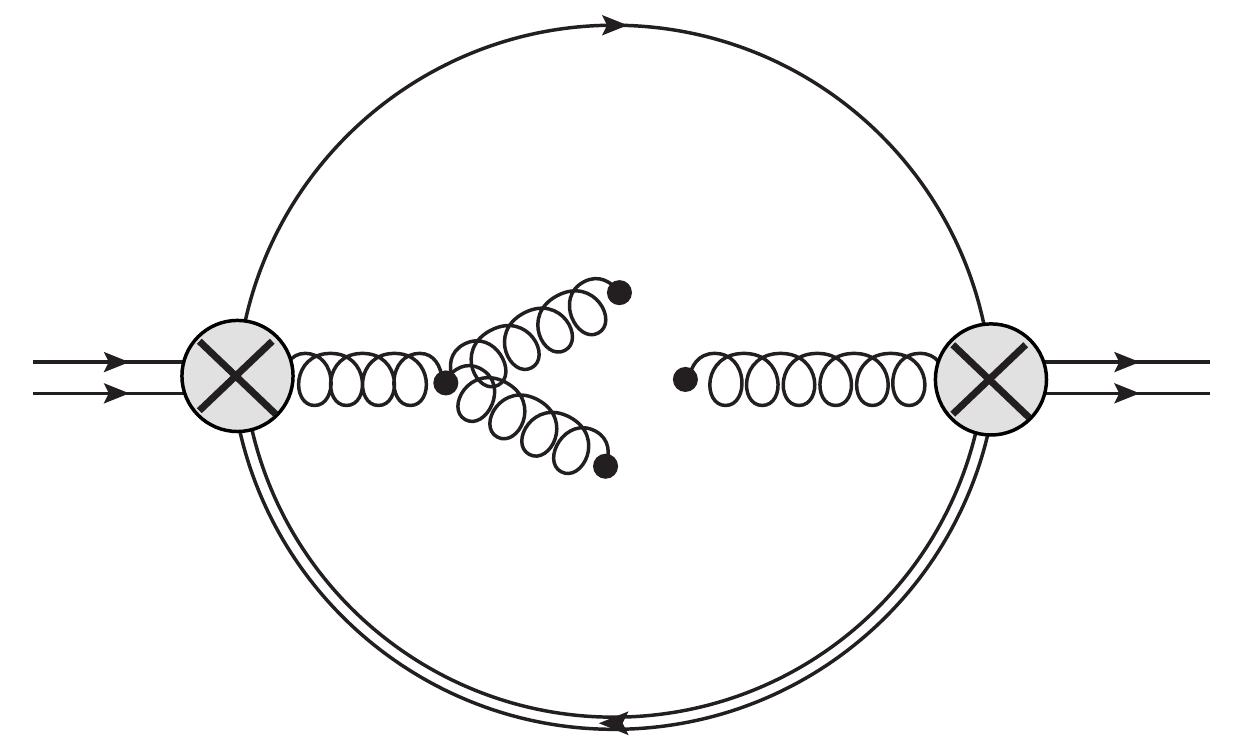}}\hfill
  \subfloat[]{%
    \includegraphics[width=.142\textwidth]{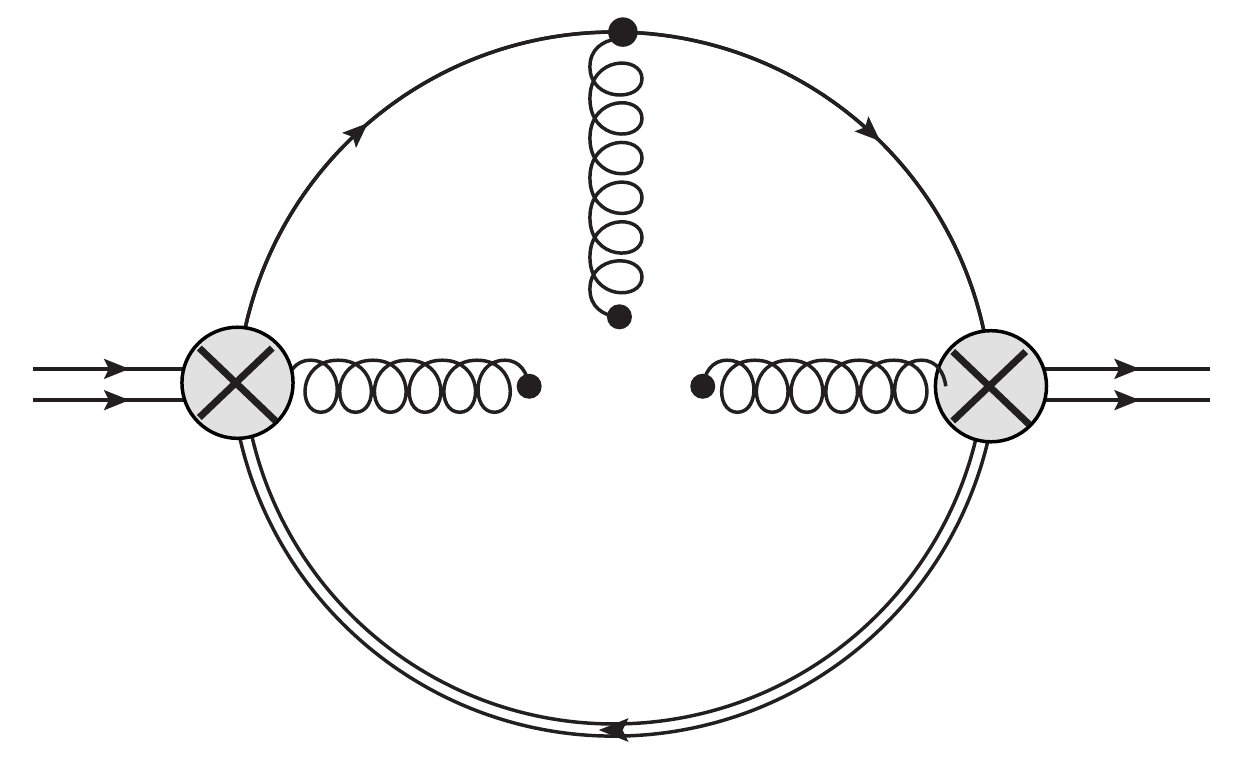}}\hfill
  \subfloat[]{%
  	\includegraphics[width=.142\textwidth]{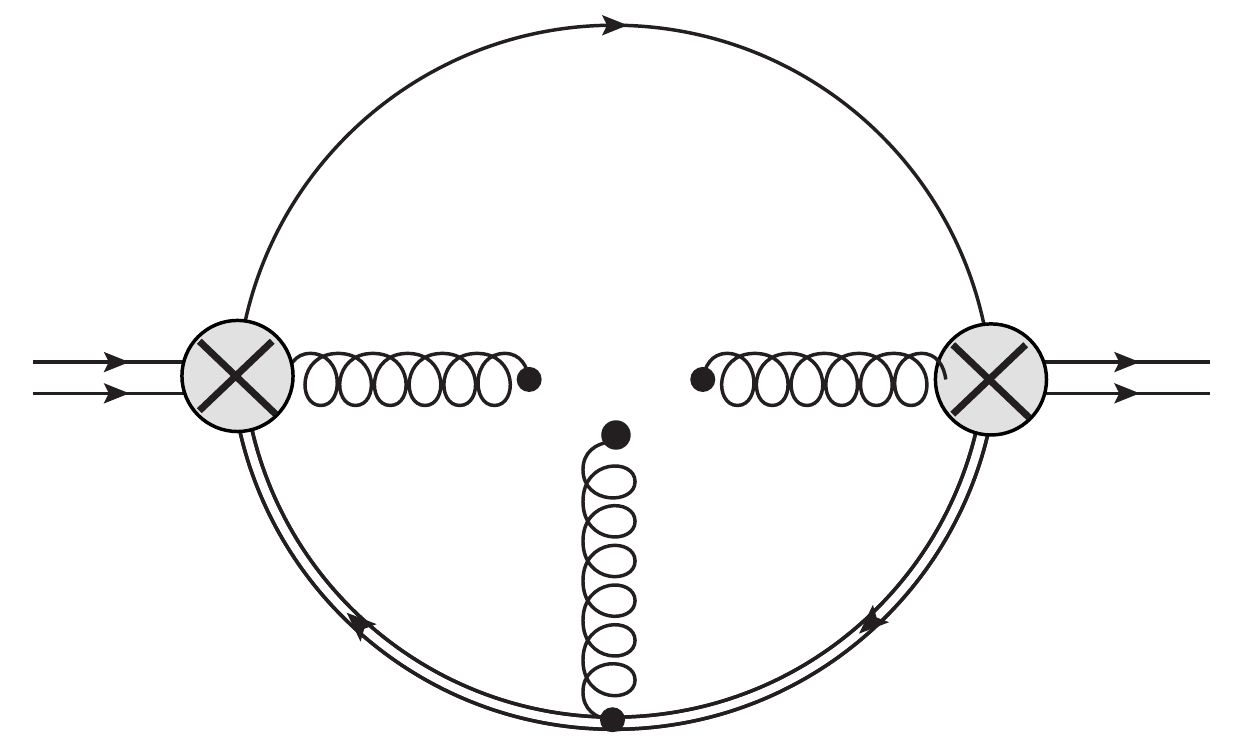}}\hfill
  \subfloat[]{%
    \includegraphics[width=.142\textwidth]{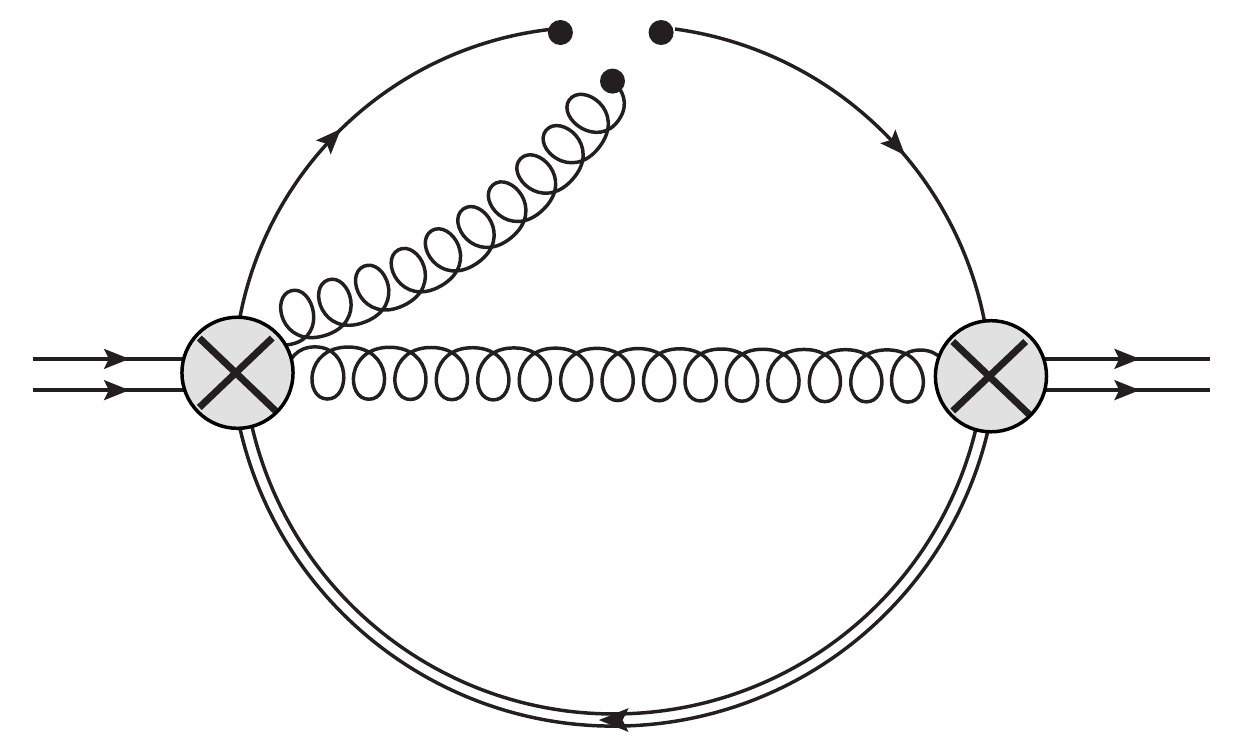}}\hfill\vspace{-0.5cm}
  \subfloat[]{%
    \includegraphics[width=.142\textwidth]{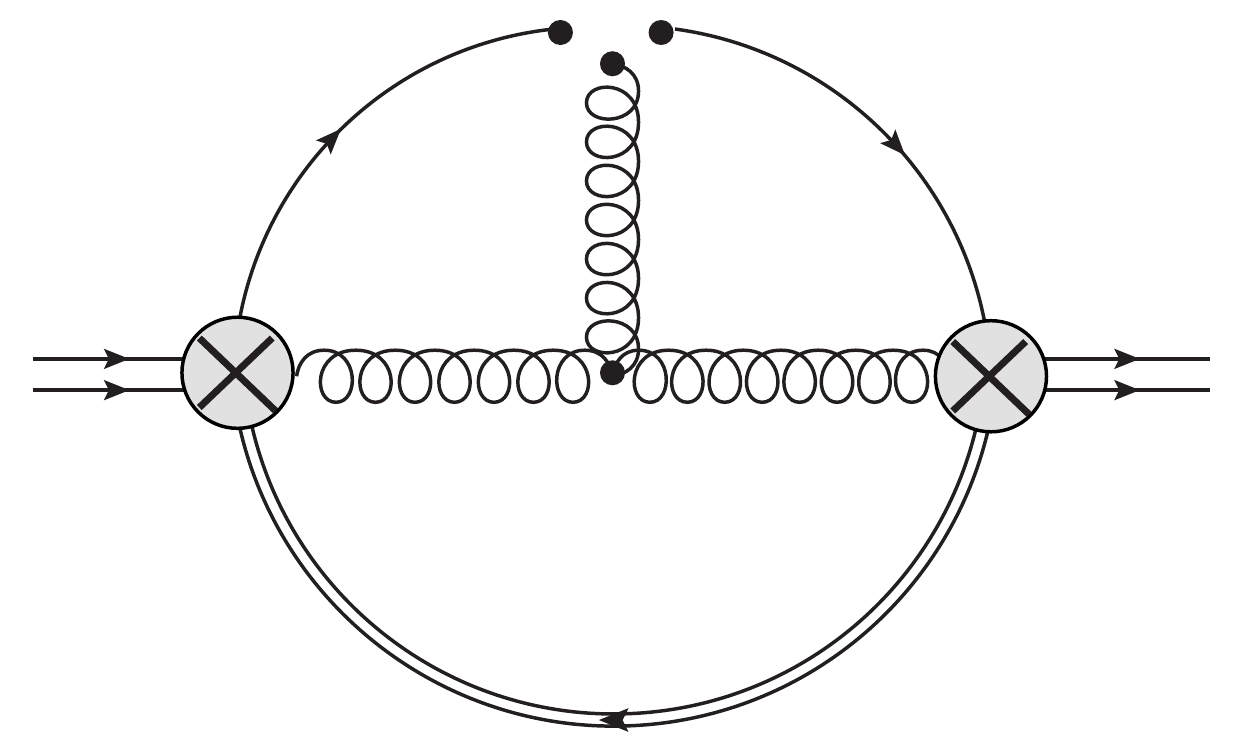}}\hfill
  \subfloat[]{%
    \includegraphics[width=.142\textwidth]{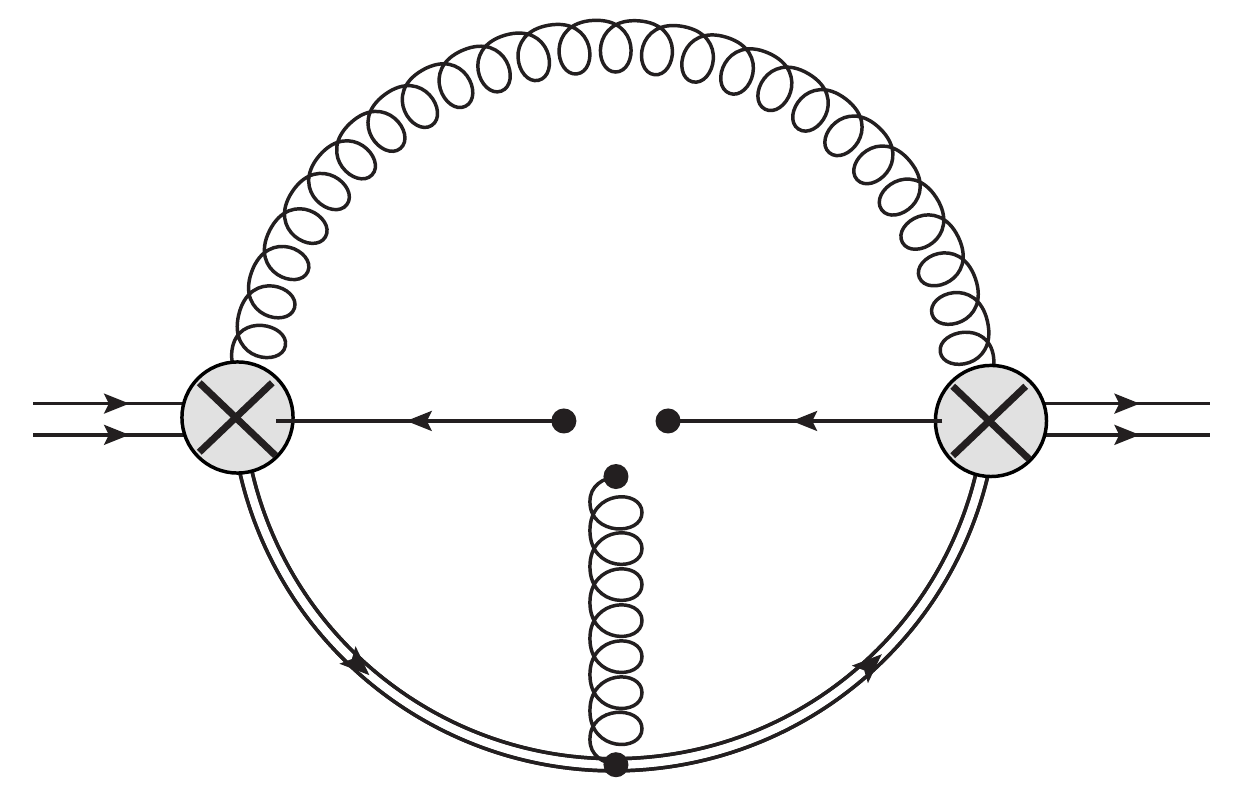}}\hfill
  \subfloat[]{%
    \includegraphics[width=.142\textwidth]{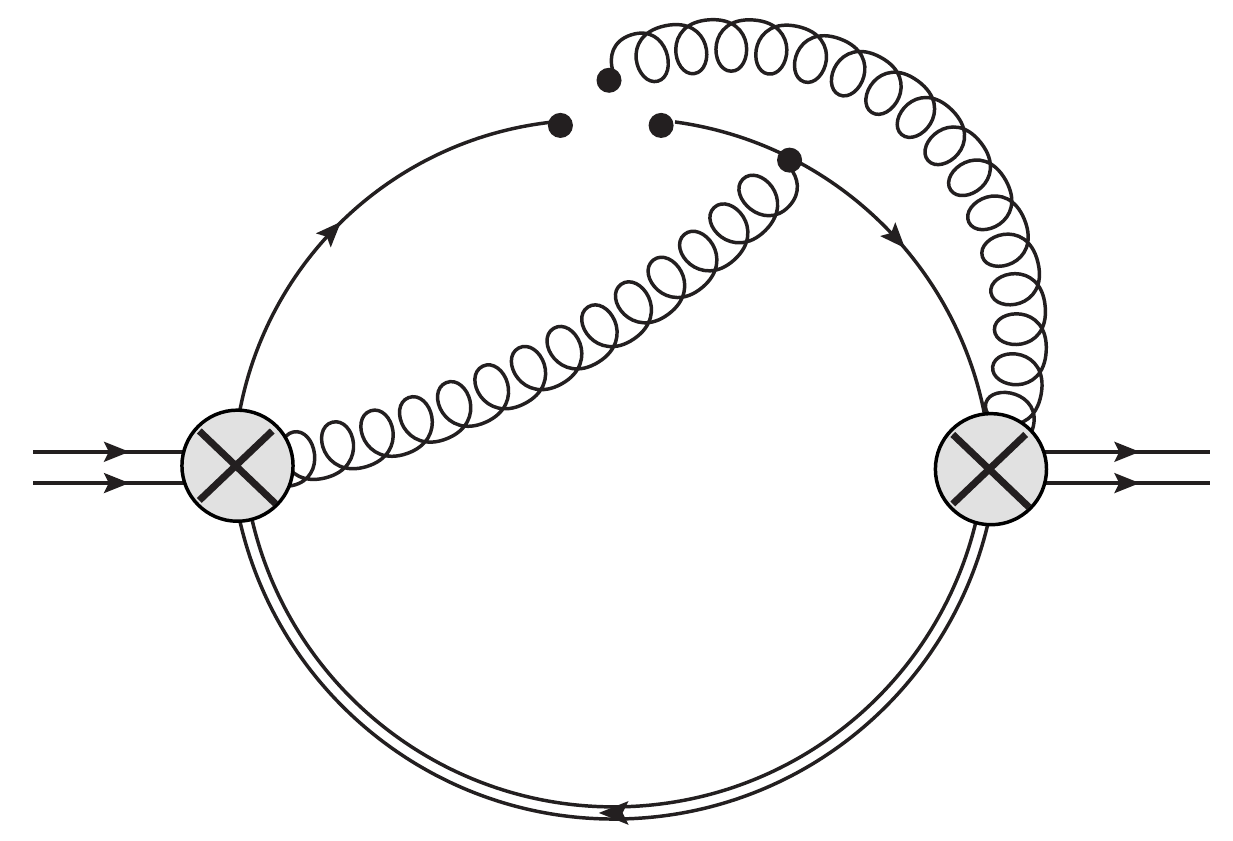}}\hfill
  \subfloat[]{%
    \includegraphics[width=.142\textwidth]{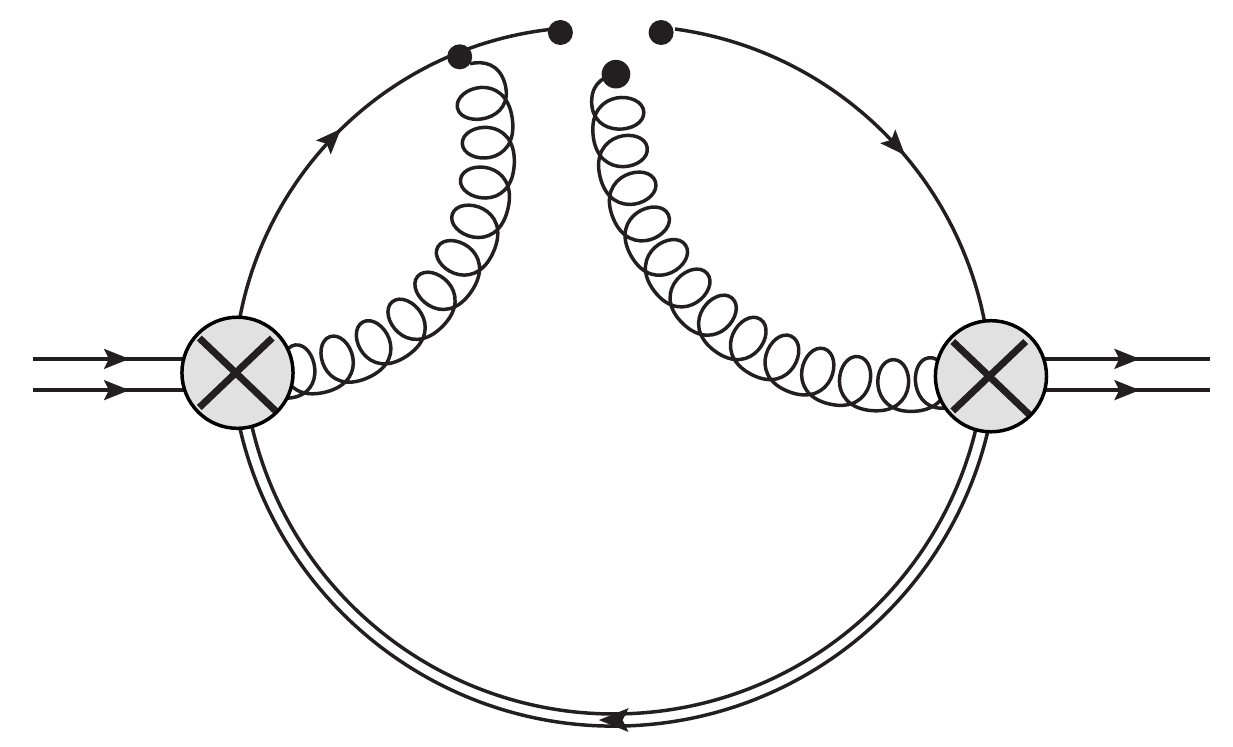}}\hfill
  \subfloat[]{%
    \includegraphics[width=.142\textwidth]{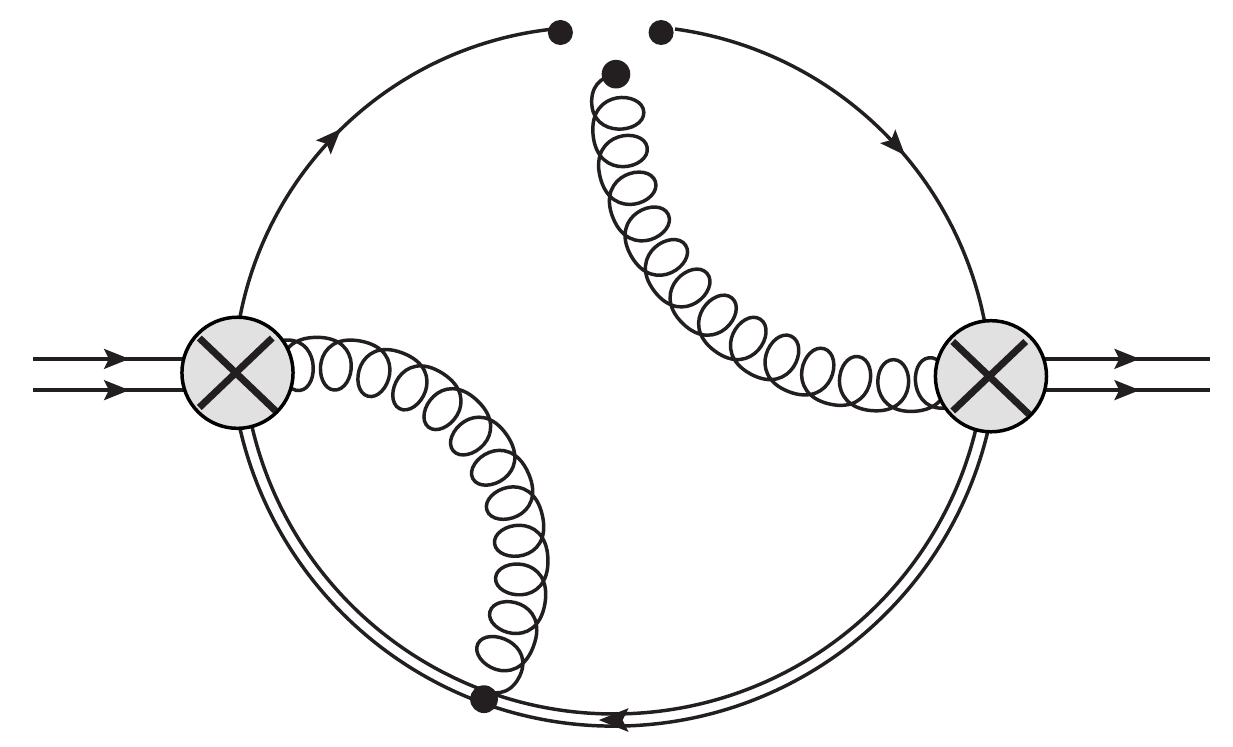}}\hfill
  \subfloat[]{%
    \includegraphics[width=.142\textwidth]{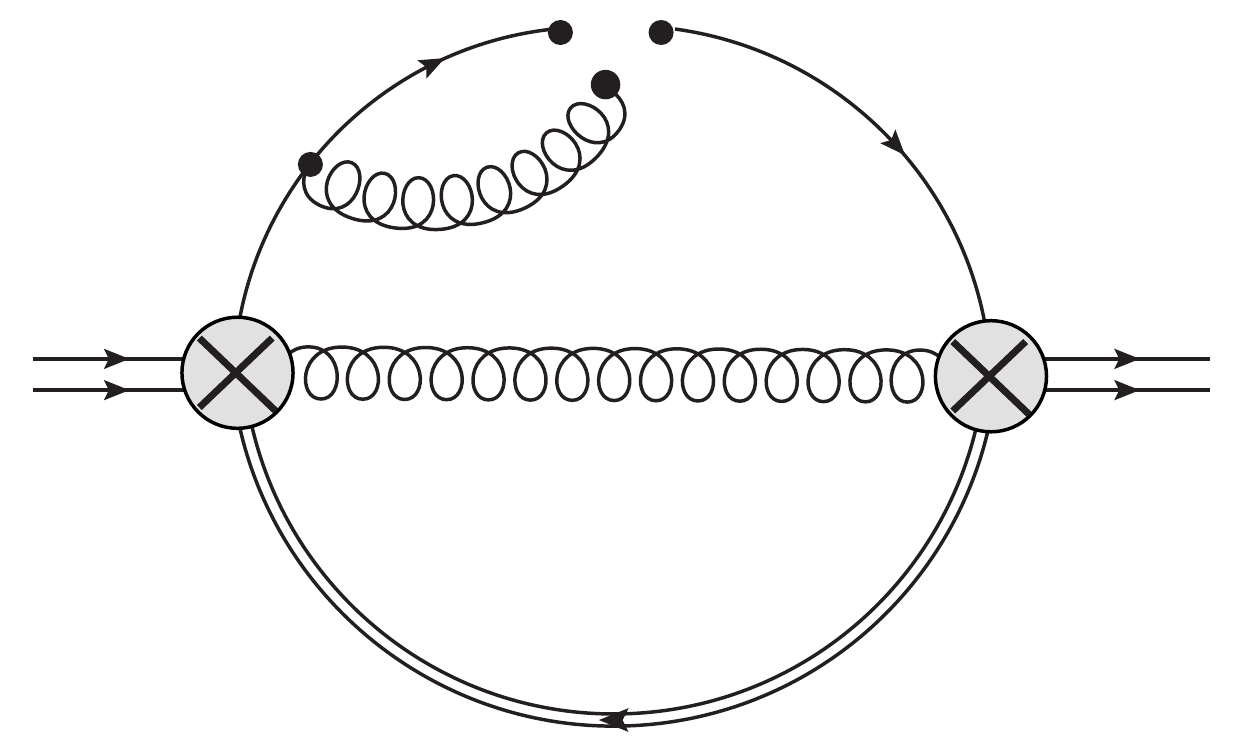}}\hfill
  \subfloat[]{%
    \includegraphics[width=.142\textwidth]{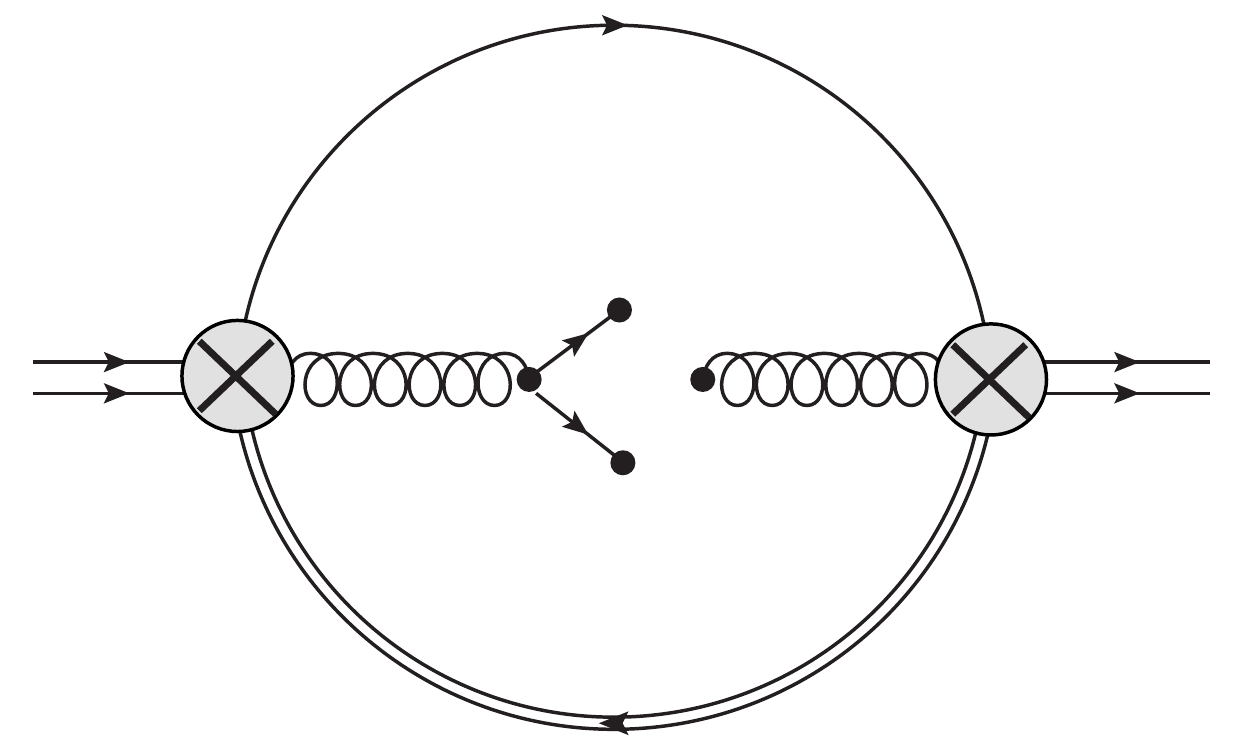}}\hfill\vspace{-0.5cm}
\captionof{figure}{The Feynman diagrams calculated for the correlator~(\protect\ref{correlator}). Single solid lines correspond to light quark propagators whereas double solid lines correspond to heavy quark propagators. All Feynman diagrams are drawn using
JaxoDraw~\protect\cite{BinosiTheussl2004}}
\label{feynman_diagrams}
\end{figure}

The Wilson coefficients are 
computed to leading-order (LO) in $g_s$
using coordinate-space, fixed-point gauge techniques 
(see \cite{BaganAhmadyEliasEtAl1994}).
Light quark masses are included in perturbation theory through a light
quark mass expansion, but have been set to zero in all other OPE terms.
The contributing Feynman diagrams are depicted in 
Figure~\ref{feynman_diagrams}.
Divergent integrals are handled using dimensional regularization 
in $D=4+2\epsilon$ spacetime dimensions at a renormalization scale $\mu^2$. We use the program TARCER~\cite{MertigScharf1998}
to reduce complicated, two-loop integrals to a small collection of 
simple basic integrals, all of which are well-known for the diagrams under consideration.
\vspace{-0.5cm}
\vspace{-0.5cm}

\section{Results of Laplace Sum-Rules Analysis}
\begin{figure}
\centering
\includegraphics[width=0.6\linewidth]{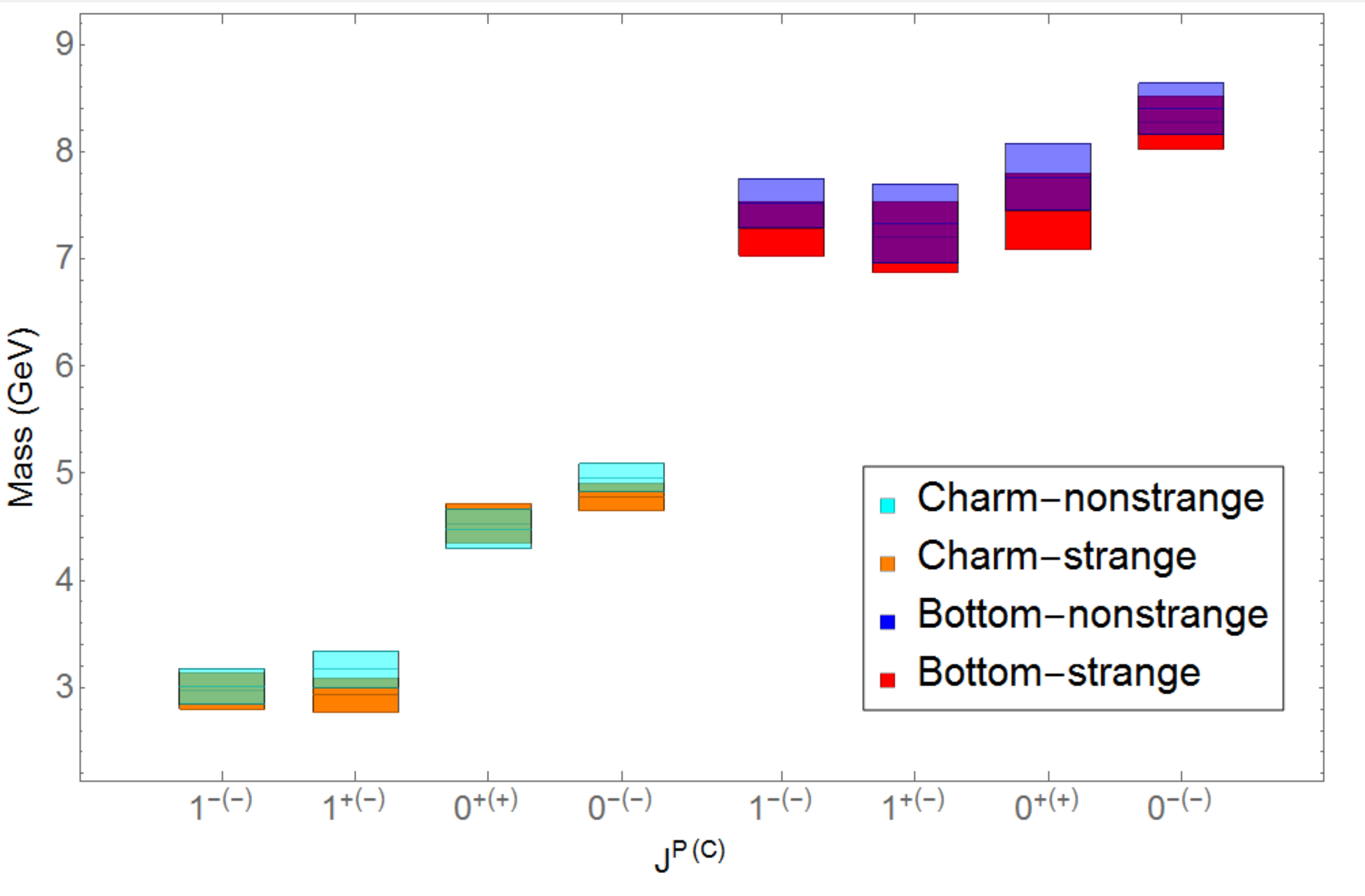}
\captionof{figure}{Summary of mass predictions with uncertainties for charm and bottom hybrid systems for the stabilizing $J^{P(C)}$ channels; channels which have been omitted do not stabilize.}\label{fig.MassSpectra}
\end{figure}

Performing a Laplace sum-rules analysis
of all eight distinct $J^{P(C)}$ combinations defined according to Table~\ref{JPC_table}, we present the mass predictions and estimated uncertainties in Figure \ref{fig.MassSpectra}. For a more in-depth discussion, please refer to \cite{HoHarnettSteele2016}.

\section{Conclusions}
Predictions for the masses of heavy-strange and heavy-nonstrange hybrid mesons for $J^{P} \in \left\{ 0^{\pm},~1^{\pm}\right\}$ are briefly presented, utilizing QCD sum-rules and improving upon the calculations of \cite{GovaertsReindersWeyers1985} by updating the non-perturbative parameters in the calculation, and including higher dimensional condensates in the OPE that have been shown important to sum-rule stability. A complete discussion of the analysis and results may be found in \cite{HoHarnettSteele2016}. A degeneracy is observed in the heavy-light and heavy-strange states, 
and stabilization in the previously unstable $0^{-(-)}$ and $1^{-(+)}$ channels \cite{GovaertsReindersWeyers1985} driven by the addition of the higher dimensional mixed and 6d gluon condensate contributions. As a consequence of these higher dimensional contributions, the $1^{+(+)}$ channel is destabilized from the original analysis of \cite{GovaertsReindersWeyers1985}.

\textbf{Acknowledgements-} We are grateful for financial support from the Natural Sciences and Engineering Research Council of Canada (NSERC).

\end{document}